\documentclass{PoS}
\usepackage{amsmath,amsthm,amssymb,latexsym} 
\usepackage{wrapfig}
\usepackage{float}
\usepackage{tikz}
\usetikzlibrary{decorations.pathmorphing}
\usetikzlibrary{decorations.pathreplacing,decorations.markings}
\pdfoutput=1

\title{Towards string breaking with 2+1 dynamical fermions using the stochastic
LapH method}

\ShortTitle{Towards string breaking with 2+1 dynamical fermions using the stochastic
LapH method}

\author{Vanessa Koch\thanks{Speaker}, John Bulava, Ben H\"orz, Mike Peardon \\
        School of Mathematics, Trinity College Dublin\\
			  Dublin 2, Ireland \\
				E-mail:\email{kochv@maths.tcd.ie, jbulava@maths.tcd.ie, hoerz@maths.tcd.ie, mjp@maths.tcd.ie}}
				
                               \author{Francesco Knechtli\\
				Dept. of Physics, University of Wuppertal\\
				Gaussstrasse 20, D-42119 Germany\\
				E-mail: \email{knechtli@physik.uni-wuppertal.de}} 
				
                                \author{Graham Moir\\
				Dept. of Physics, University of Wuppertal\\
				Gaussstrasse 20, D-42119 Germany\\
				Dept. of Applied Mathematics and Theoretical Physics, University of Cambridge\\
                                Wilberforce Road, Cambridge, CB3 0WA, UK\\
				E-mail: \email{graham.moir@damtp.cam.ac.uk}}

                                \author{Colin Morningstar\\
                                Dept. of Physics, Carnegie Mellon University\\
				Pittsburgh, PA 15213, USA\\
                                E-mail: \email{colin\_morningstar@cmu.edu}}

\abstract{We investigate the use of stochastically estimated light quark propagators in correlation functions involving a static color source. To this end we compute the static-light meson pseudoscalar correlation function in the stochastic LapH framework, using an ensemble of $N_f= 2+1$ gauge configurations generated through the CLS effort. We extract the static-light as well as the static-strange mass with good statistical precision. Together with the static potential, we obtain a preliminary estimate for the expected mixing region.}

\FullConference{The 33rd International Symposium on Lattice Field Theory\\
                 14 -18 July  2015\\
                 Kobe International Conference Center, Kobe, Japan}

\begin{document}

\section{Introduction}
String breaking, the transition of the static quark anti-quark string into a static-light meson-antimeson system, provides an intuitive example of a strong decay and one of the defining characteristics of a confining gauge theory. 
It has been thoroughly investigated for the SU(2) Higgs model, e.g. \cite{knechtli1,knechtli2,philipsen}. For QCD only an exploratory $N_f= 2$ study for one lattice size, one lattice spacing and one relatively heavy quark mass exists \cite{bali}.

We want to investigate string breaking using distillation and the stochastic LapH method to 
get better statistical accuracy. Furthermore we want to examine the quark mass dependence as well as the effect of a third sea-quark flavor, which is expected to result in a second mixing-phenomenon due to the formation of a strange-antistrange pair. The first necessary step is to test stochastically estimated light quark propagators in correlation functions involving a static color source. To that end we calculate $E_{stat}$, the mass of the static-light meson. 

An estimate of the string breaking distance $r_b$ can be calculated utilizing $E_{stat}$ and the static potential $V (r)$.  To observe the phenomenon the lattice has to have a size $L > 2r_b$.

The measurements are done on a subset of evenly-spaced configurations of the N200 ensemble with $N_f = 2 + 1$ flavors of non-perturbatively  $O(a)$-improved Wilson fermions generated by the CLS effort \cite{cls}. The lattice-size is $N_t \times N_s^3 = 128 \times 48^3$ with an estimated  isotropic lattice spacing of $a\approx$ 0.064 fm and pion and kaon mass of $m_{\pi}\approx$ 280 MeV and $m_{K}\approx$ 460 MeV respectively.
Due to the open temporal boundary conditions translation invariance in time is broken, however these boundary effects decrease exponentially \cite{Luscher}. In order to ensure that boundary effects are absent, we use a single source time at $t_0 = 32$, for further reference see \cite{Ben}.

\section{The stochastic Laph method}
Using smeared quark fields is a well-established way to reduce high frequency modes and thus especially important if 
one wants to extract low-lying energies.
A special quark-field smearing algorithm is distillation \cite{mike}, it applies a low-rank operator to define smooth
fields that are to be used in hadron creation operators.

A quark line ${\cal D}$ will be the following product of matrices after applying the Distillation operator to each quark field:

\begin{equation}
{\cal D}=S \Omega^{-1} S =V(V^{\dagger}\Omega^{-1}V)V^{\dagger},
\end{equation}
where $\Omega^{-1}=\gamma_4D$, which is conventional to ensure hermiticity for some correlation matrices, and 

\begin{equation}
S=VV^{\dagger}, \ \ \ \ V^{\dagger}V=1 ,
\end{equation}

is the projection operator into a smaller subspace, spanned by the eigenvectors of the Laplacian. 
Instead of computing and storing all elements of $ \Omega^{-1}$, one has to find the smaller matrix $V^\dagger \Omega^{-1} V$. Only $N_v N_t N_d$
inversions are required for each quark mass and gauge configuration in the ensemble where $N_d=4$ denotes the number of Dirac spin components and $N_v$ the number of eigenvectors.
The problem is that the number of eigenvectors $N_v$ needed for a fixed smearing cutoff parameter $\sigma^2$, scales linearly with the volume of the lattice. Thus the number of inversions needed will still be too high to be feasible for bigger lattices. A possible solution is to stochastically estimate the matrix $V^\dagger \Omega^{-1} V$ using random noise vectors to facilitate all-to-all propagation even for bigger lattices \cite{morningstar}.
Since the path integrals are being evaluated using a Monte Carlo based method, the statistical errors for the hadron correlators are
limited by the statistical fluctuations arising from gauge-field sampling. This means that the quark lines only have to be estimated to
a comparable accuracy, exact treatment is not necessary.

Noise vectors $\rho$ can be introduced in the smaller subspace spanned by the eigenvectors of the Laplacian, the so-called LapH-subspace. These vectors have spin, time and eigenmode  indices, each component of $\rho$ is a random $Z_4$ variable and  $E(\rho)=0$ and $E(\rho\rho^\dagger)=\mathbb I$. $E$ denotes the expectation value. 
The problem is that the variances of the stochastic estimates are usually much too large, the noisy estimates need 
variance reduction techniques to separate signal from noise. This is possible through dilution of the noise vectors \cite{dilution}.
The dilution projectors $P^{(b)}$ are products of time dilution, spin dilution, and LapH eigenvector dilution projectors. A propagator can then be stochastically estimated:
\begin{eqnarray}
 {\cal D}  &=&   S \ \Omega^{-1} V V^\dagger ,\nonumber\\
  &=& \textstyle\sum_b  S \ \Omega^{-1} V P^{(b)}P^{(b)\dagger} 
 V^\dagger ,\nonumber\\
 &=& \textstyle\sum_b   S \ \Omega^{-1} VP^{(b)}E(\rho\rho^\dagger)
  P^{(b)\dagger}  V^\dagger ,\nonumber\\
 &=& \textstyle\sum_b E\Bigl( \!   S\  \Omega^{-1} VP^{(b)}\rho
     \,( V P^{(b)}  \rho)^\dagger \!\Bigr),
 \end{eqnarray}
The estimate of a quark line over a given gauge configuration now reads: 
\begin{eqnarray}
{\cal D}_{uv}^{(ij)} \approx \frac{1}{N_R}\delta_{ij}\textstyle\sum_{r=1}^{N_R}\sum_{b=1}^{N_b}  \varphi^{r[b]}_u \varrho^{r[b]\ast}_v . 
\end{eqnarray}
The subscripts $u$, $v$ are compound indices indicating space, time, color, spin and $i$, $j$ denote the flavor of the source and sink field, and  
we have defined smeared-diluted quark source and quark sink vectors as
\begin{eqnarray}
 \varrho^{r[b]} =   V P^{(b)}\rho^r, \ \ \ \ \ \varphi^{r[b]} =  S \ \Omega^{-1}\ V P^{(b)}\rho^r.
\end{eqnarray}

%

\section{Static-light meson}

The correlation function for a static-light quark, with interpolator given by $\mathcal{O} \ \ \equiv \ \ \overline{Q} \gamma_5 q^i$ , 
where i is the quark flavor, is given by:

\begin{wrapfigure}{l}{0.1\textwidth}
\begin{center}
\begin{tikzpicture}[scale=0.5]
\draw[gray, thick, decorate, decoration={snake, segment length=2.6mm, amplitude=0.8mm}] (-2,0) .. controls (-1,2).. (-2,4);
\filldraw[black] (-2,4)  node[anchor=north, font=\footnotesize] {$\overline{Q}(\mathbf{x},t)\hspace{18mm}    q(\mathbf{x},t)$};
\filldraw[black] (-2,0)  node[anchor=south, font=\footnotesize] {$Q(\mathbf{x},0)\hspace{18mm}     \overline{q}(\mathbf{x},0)$};
\draw[gray, thick, decoration={markings, mark=at position 0.5 with {\arrow{<}}},postaction={decorate}] (-2,0) -- (-2,4);
\end{tikzpicture}
\end{center}
\end{wrapfigure}
\begin{align}
C(t)=   \langle [\overline{Q}(\mathbf{x},t)\gamma_5 q^i(\mathbf{x},t) \overline{q}^i(\mathbf{x},0)\gamma_5 Q(\mathbf{x},0)]\rangle \\ \nonumber
=  \langle tr(\gamma_5 \underbrace{D(t,0)\gamma_4}_{\text{light  propagator}}\gamma_4 \gamma_5  \underbrace{\mathcal{P}(0,t)P_-}_{\text{static  propagator}}  )\rangle \\ \nonumber
=  \langle tr((\gamma_5)^2 \Omega^{-1}\gamma_4 \mathcal{P}(0,t)P_+ )\rangle
\end{align}

$\mathcal{P}(0,t)$ is a timelike Wilson-line of HYP-smeared \cite{hyp} links from $(x,t)$ to $(x,0)$, the projector $P_{\pm}$ is given by
\begin{equation}
 P_\pm={\frac{1\pm\gamma_4}{2}} .
\end{equation}
The static propagator we use is a modification of the static propagator derived by Eichten and Hill, in \cite{staticdellamorte} it was proven that using actions with HYP-smeared links improves the signal-to-noise ratio at large euclidean times. 

If the light quark line is stochastically estimated according to the stochastic LapH method, the correlation function takes the following form:
\begin{align}
\begin{split}
C(t) &\approx \langle tr( \underbrace{[ S\Omega^{-1}VP^{(b)}\rho^r]}_{\text{sink}}\underbrace{(VP^{(b)}\rho^r)^{\dagger}}_{\text{source}}\gamma_4 \mathcal{P}(0,t)P_+)\rangle_{MC}\ \\ 
&=\langle tr( \varphi^{r[b]} \varrho^{r[b]*} \cdot (\gamma_4 \mathcal{P}(0,t) P_+ ) )\rangle_{MC} \\ 
&=\langle   \varrho^{r[b]*} \cdot (\gamma_4 \mathcal{P}(0,t) P_+ \varphi^{r[b]}) \rangle_{MC} , 
\end{split}
\end{align}

Now we have to compute just a scalar product of source and sink with an inserted timelike Wilson-line, projected onto upper spin components.

We use $N_R=4$ and $N_R=2$ noise sources for light and strange quarks respectively. Using $N_b=32$ dilution projectors - full time and spin dilution
with interlace-8 LapH eigenvector dilution - requires a total of $128$ light quark inversions and 
$64$ strange quark inversions per configuration. The inversions need to be done once and can be reused for other calculations. Only the projected sinks, complex numbers, are stored. This reduces the needed storage and amounts to roughly  300MB per configuration for the light and strange quarks combined. Even for more complex calculations that need more inversions, the amount of storage will be feasible.

Our results for the static-light and static-strange mass show a good accuracy, even though the first results were obtained on a small number of configurations.
The individual points are obtained through a correlated fit of the correlation function to a single exponential within the interval $[t_{min},t_{max}=29]$ using Bootstrap error estimation. The covariance matrix was estimated on the original data and kept 'frozen' for all Bootstrap samples. The fitted mass is calculated as a function of $t_{min}$.

\begin{figure}[H]
\centering
\includegraphics[width=0.7\linewidth]{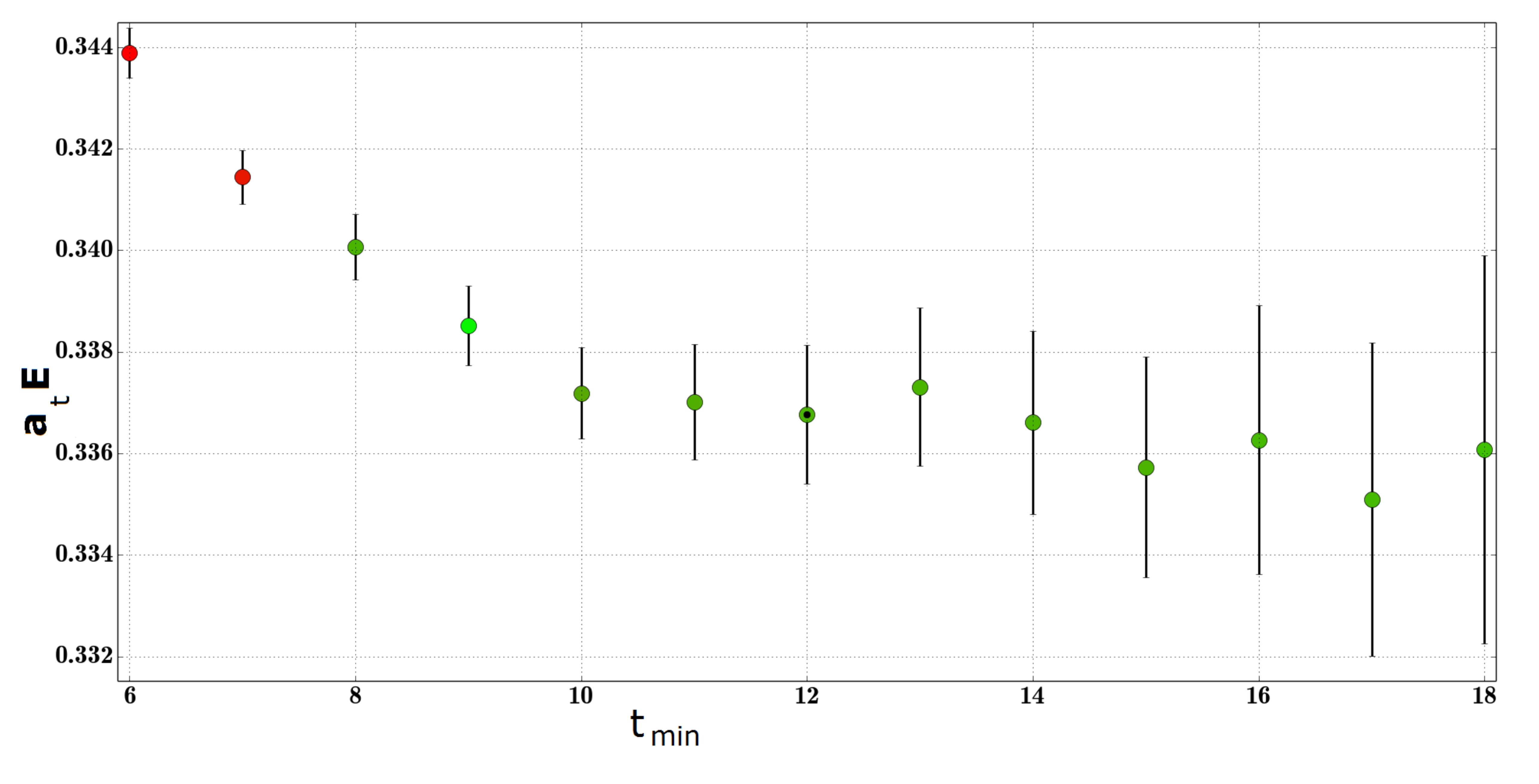}
\caption{$t_{min}$-plot for the static-light meson on 200 configurations. The color of the dots indicates if $\chi^2$ is acceptable, where for green dots $\chi^2/\text{d.o.f.}\approx1$. The black dot represents the chosen value with a relative uncertainty of $0.4\%$ .}
\end{figure}

\begin{figure}[H]
\centering
\includegraphics[width=0.7\linewidth]{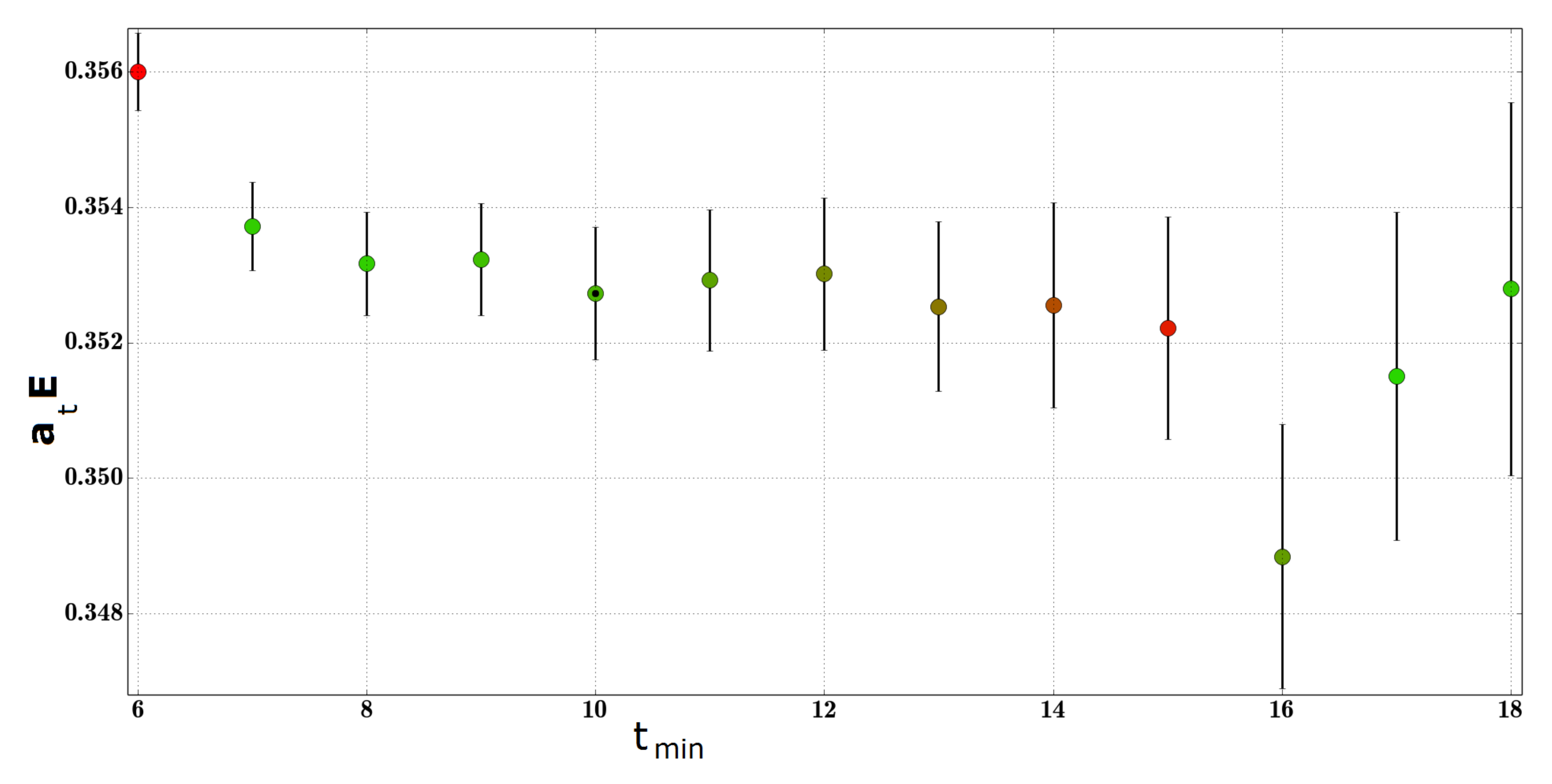}
\caption{$t_{min}$-plot for the static-strange meson, on 100 configurations. The color of the dots indicates if $\chi^2$ is acceptable, where for green dots $\chi^2/\text{d.o.f.}\approx1$. The black dot represents the chosen value with a relative uncertainty of $0.3\%$. }
\end{figure}

\section{Static Potential}
To get an estimate for the string breaking distance on this ensemble, we employ rectangular on-axis Wilson-loops $W(T,r)$ to serve as an observable for the static potential:
\begin{equation}
 \langle W(T,r)\rangle \stackrel{T \rightarrow \infty}{\propto} e^{-V(r)T} , 
\end{equation}
Following the method presented in \cite{wloops}, using Leder's Wilson loop package \cite{bjorn}, we first smear all gauge-links using HYP2 parameters \cite{hyp2} and then construct a variational basis using HYP smeared spatial links. The generalized eigenvalue problem using this basis is solved to extract the static potential $V(r)$.
$V(r)$ now is renormalized by subtracting twice the energy of a static-light meson.

\begin{figure}
\centering
\includegraphics[width=0.8\linewidth]{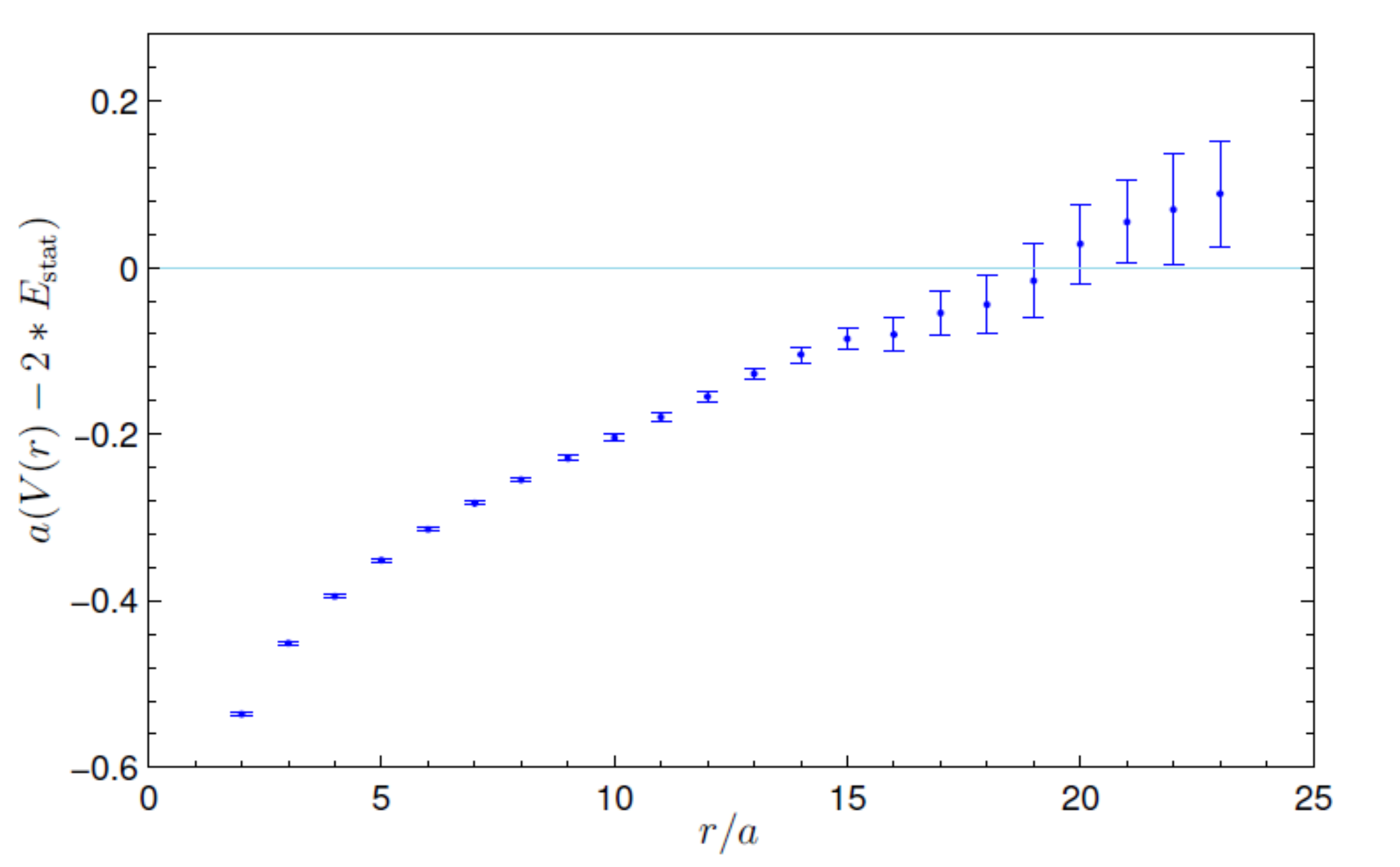}
\caption{Ground state potential $V(r)$ on 100 configurations}
\end{figure}

For the Sommer-parameter \cite{sommer} defined through the static force $F(r)=V'(r)$ as the solution of 
\begin{equation}
 r^2F(r)|_{r=r_0}=1.65                  
\end{equation}
we find 
\begin{equation}
 r_0 = 7.77(22) a \approx 0.5 fm 
\end{equation}


\section{Results and outlook}
Using the stochastic LapH method allows for accurate determinations of temporal correlations for static-light mesons.

String breaking is expected to occur as soon as $[V(r) - 2E_{stat} ] > 0$, for the N200 we find  the following value for the expected string breaking distance:
      \begin{equation}
     r_b \approx 19 a \quad \quad\frac{r_b}{r_0} \approx 2.4  .  
      \end{equation}
This means that it is possible to resolve string breaking on this lattice, according to the requirement $L > 2r_b$.\newline
The result is particularly interesting in comparison with the results of Bali et al \cite{bali}, who found $\frac{r_b}{r_0} \approx 2.5$ with $m_{\pi}\approx 640 MeV$ and $N_f=2$.

 The next step has to be a full analysis of string breaking including mixing effects using the following ansatz for the correlation function:

\begin{eqnarray}
C(t)\quad=\left(
\begin{array}{cc}C_{QQ}(t)&C_{Qq}(t)\\
C_{qQ}(t)&C_{qq}(t)\end{array}\right) 
\propto\left(\begin{array} {rl}
\begin{tikzpicture}[baseline={([yshift=-.8ex]current bounding box.center)}]
\draw[gray, thick] (0,0) -- (0,1);
\draw[gray, thick] (0,1) -- (1,1);
\draw[gray, thick] (1,1) -- (1,0);
\draw[gray, thick] (1,0) -- (0,0);
\end{tikzpicture}
&\quad \ \ \  \begin{tikzpicture}[baseline={([yshift=-.8ex]current bounding box.center)}]
\draw[gray, thick] (0,0) -- (0,1);
\draw [decorate, decoration={snake, segment length=1.5mm, amplitude=0.5mm}](0,1) -- (1,1);
\draw[gray, thick] (1,1) -- (1,0);
\draw[gray, thick] (1,0) -- (0,0);
\end{tikzpicture} \\&\\
 \begin{tikzpicture}[baseline={([yshift=-.8ex]current bounding box.center)}]
\draw[gray, thick] (0,0) -- (0,1);
\draw[gray, thick] (0,1) -- (1,1);
\draw [gray, thick](1,1) -- (1,0);
\draw[decorate, decoration={snake, segment length=1.5mm, amplitude=0.5mm}](1,0) -- (0,0);
\end{tikzpicture}
&\quad -\begin{tikzpicture}[baseline={([yshift=-.8ex]current bounding box.center)}]
\draw [gray, thick](0,0) -- (0,1);
\draw [decorate, decoration={snake, segment length=1.5mm, amplitude=0.5mm}](0,1) -- (1,1);
\draw [gray, thick](1,1) -- (1,0);
\draw [decorate, decoration={snake, segment length=1.5mm, amplitude=0.5mm}](1,0) -- (0,0);
\end{tikzpicture}+\begin{tikzpicture}[baseline={([yshift=-.8ex]current bounding box.center)}]
\draw [gray, thick](0,0) -- (0,1);
\draw (0,0) .. controls (0.2,0.5) .. (0,1)[decorate, decoration={snake, segment length=1.5mm, amplitude=0.5mm}];
\draw [gray, thick](1,1) -- (1,0);
\draw (1,0) .. controls (0.8,0.5) .. (1,1)[decorate, decoration={snake, segment length=1.5mm, amplitude=0.5mm}];
\end{tikzpicture} 
\end{array}\right),
\end{eqnarray}

where the diagonal entries correspond to a string and a two meson state.

Furthermore we plan to repeat the calculation on other CLS ensembles to study the dependence of the string breaking distance on different quark masses.

To investigate the effect of including the strange quark into the calculation, the correlation matrix will have to be expanded to a $3 \times 3$ matrix.

\section*{Acknowledgements}

The authors wish to acknowledge the DJEI/DES/SFI/HEA Irish Centre for High-End Computing (ICHEC) for the provision of computational facilities and support.
The computations for the static potential were performed on the cluster 'stromboli' at the university of Wuppertal.
The code for the calculations using the stochastic LapH method is built on the USQCD QDP++/Chroma library \cite{chroma}.
We acknowledge PRACE for awarding CLS access to resource SuperMUC based in Germany at LRZ, Munich. CJM acknowledges support from the U.S.~NSF under award PHY-1306805.
BH and VK are supported by Science Foundation Ireland under Grant No. 11/RFP/PHY3218 and No. 11/RFP.1/PHY/3201 respectively.

\providecommand{\href}[2]{#2}\begingroup\raggedright\endgroup

\end{document}